\documentclass[]{aastex631}

\begin{document}

\title{White dwarf -- red giant star binaries as Type Ia supernova progenitors: with and without magnetic confinement}



\author[:0000-0001-7003-4220]{Iminhaji Ablimit}
\affiliation{CAS Key Laboratory for Optical Astronomy, National Astronomical Observatories, Chinese Academy of Sciences, Beijing 100012, China}
\affiliation{Department of Astronomy, Kyoto University, Kitashirakawa-Oiwake-cho, Sakyo-ku, Kyoto 606-8502, Japan}

\author{Philipp Podsiadlowski}
\affiliation{University of Oxford, St Edmund Hall, Oxford OX1 4AR, United Kingdom}

\author{Rosanne Di Stefano}
\affiliation{Institute for Theory and Computation, Center for Astrophysics, Harvard University \& Smithsonian, 60 Garden Street, Cambridge, MA 02138, USA}

\author{Saul A. Rappaport}
\affiliation{M.I.T., Department of Physics and Kavli Institute for Astrophysics and Space Research, 70 Vassar St., Cambridge, MA, 02139, USA}

\author{James Wicker}
\affiliation{National Astronomical Observatories, Chinese Academy of Sciences, Beijing 100012, China}


\correspondingauthor{Iminhaji Ablimit}
\email{iminhaji@nao.cas.cn}

\begin{abstract}

Various white-dwarf (WD) binary scenarios have been proposed trying to understand the nature and the diversity of Type Ia supernovae (SNe Ia). In this work, we study the evolution of
carbon-oxygen WD -- red giant (RG) binaries (including the role of magnetic confinement) as possible SN Ia progenitors (the so-called symbiotic progenitor channel). Using the \textsc{mesa} stellar evolution code, we calculate the time dependence
of the structure of the RG star, the wind mass loss, the Roche-lobe-overflow (RLOF)
mass-transfer rate, the polar mass-accretion rate (in the case of magnetic confinement), and the orbital and angular-momentum evolution.  We consider cases where the WD is non-magnetic and cases where the magnetic field is strong enough to force accretion onto the
two small polar caps of the WD. Confined accretion onto a small area allows for
more efficient hydrogen burning, potentially suppressing nova outbursts. This makes it easier for the WD to grow in mass towards the Chandrasekhar mass limit and explode
as a SN Ia. With magnetic confinement, the initial parameter space of the
symbiotic channel for SNe Ia is shifted towards shorter orbital periods and lower donor masses compared to the case without magnetic confinement. Searches for low-mass
He WDs or relatively low-mass giants with partially stripped envelopes that survived
the supernova explosion and are found in SN remnants will provide crucial insights for our understanding of the contribution of this symbiotic channel.

\end{abstract}

\keywords{(stars:) binaries (including multiple): close -- stars: evolution -- (stars:) white dwarfs -- stars: magnetic field -- (stars:) supernovae: general - stars: late-type}

\section{Introduction}

Type Ia supernovae (SNe Ia) are known as one of the very energetic astronomical phenomena which are a key for understanding the evolution of the Universe; they have generally been considered the results of thermonuclear explosions of carbon-oxygen white dwarfs (CO WDs; Hoyle \& Fowler 1960) in close binary systems.
In both observational and theoretical studies, it is still hard to arrive at a robust conclusion on the binary stellar evolution pathways that lead to SNe Ia (for recent reviews see e.g., Maoz, Mannucci \& Nelemans 2014; Maeda \& Terada 2016; Livio \& Mazzali 2018; Soker 2019; Jha et al.\ 2019), and indeed many different possible routes could contribute to the SN Ia population (see, e.g., Soker [2018] for a
comparison of the different evolutionary scenarios).
The SN Ia progenitor scenarios can be summarized as follows:
(a) The single-degenerate (SD) scenario in which 
the CO WD accretes matter from a non-degenerate
companion star and grows in mass towards the Chandrasekhar mass
(e.g.\ Whelan \& Iben 1973; Nomoto 1982; Li \& van den Heuvel 1997; Langer et al. 2000; Han \& Podsiadlowski 2004;
Podsiadlowski et al. 2008; L$\ddot{\rm u}$
et al. 2009; Ruiter et al. 2009; Di Stefano \& Kilic 2012; Wang \& Han 2012; Nelemans et al. 2013; Claeys et al. 2014; Ablimit et al. 2014; Ablimit \& Maeda 2019a,b, Liu et al. 2021; Ablimit 2022).
(b) The double-degenerate (DD) scenario which contains two WDs (of which at least one is a CO WD) which will experience a merger  (e.g.\ Iben
\& Tutukov 1984; Webbink 1984;  Marsh et al. 1995; Fryer et al. 2010; Toonen et al. 2012; Pakmor et al. 2013; Sato et al. 2015; Ablimit et al. 2016; Perets et al. 2019).
(c) The  WD-WD  collision (WWC) scenario where two CO WDs collide head-on at about their free fall velocity which causes nuclear ignition at the interface (e.g., Raskin et al. 2009; Kushnir et al. 2013)
(d) The core-degenerate (CD) scenario where a CO WD companion
merges with the CO (or possibly HeCO) core of a massive asymptotic giant branch (AGB) star during a common-envelope (CE) phase (e.g., Kashi \& Soker 2011; Soker 2011;
Soker 2014). 
(e) The core-merger-detonation (CMD) scenario, where the merger
of a CO WD with the He core of a non-degenerate evolved star
during a CE phase induces a double detonation inside a common envelope (Ablimit 2021).

The initial ignition of the WD that leads to its explosion can occur in two main different ways: one is the delayed core detonation of a Chandrasekhar-mass WD (e.g., Khokhlov 1991;
Roepke \& Niemeyer 2007; Roepke et al.\ 2007; Kasen et al. 2009; Hristov et al. 2018 Lach et al.\ 2021); the other
is a helium-shell detonation near the WD surface that subsequently 
leads to a carbon detonation in the core of a sub-Chandrasekhar-mass WD (e.g., Woosley \& Weaver 1994; Shen et al. 2021), commonly referred to as a double detonation. The direct core
ignition of a WD can happen in the SD model (with the WD $+$ main-sequence (MS) star channel, WD $+$ red giant (RG) channel etc.), the DD model (with the merger of two CO WDs etc.) and the CD model,
while the double detonation occurs in some variants of the SD model (with the WD $+$ non-degenerate helium (He) star channel), the DD model (with collisions or mergers between a CO WD and a He WD or a hybrid HeCO WD) and the CMD model. In the case of a head-on collision (the WWC scenario), the explosion is initiated by a detonation either in the shocked region or in the contact region near the WD interface (depending on the mass of the WDs; Kushnir et al.\ 2013). It is still hard to constrain the detailed explosion mechanism and the progenitor models, not only because of the complexities and speculative aspects of the theoretical studies, but also because of the many intrinsic variations in observed SN Ia properties.

Observationally, photometric and spectroscopic properties of SNe Ia provide
promising clues for understanding SN Ia progenitor systems and the explosion physics.
SN Ia lightcurves observed by recent telescopes such as
the Kepler spacecraft (e.g., Dimitriadis et al. 2019; Shappee et al. 2019)
and the Transiting Exoplanet Survey Satellite (e.g., Fausnaugh et al. 2019)
have been used to search for better constraints. However,
the interpretation of these observational results is still not clear enough to allow
robust conclusions (e.g., Piro \& Nakar 2013; Magee et al. 2018; Stritzinger et al. 2018; Polin et al. 2019).
Tiwari et al.\ (2022) argued that observations of the late-time lightcurve of SN 2015F are only consistent
with a sub-Chandrasekhar-mass WD progenitor, while observations of four other events (SN 2011fe,
SN 2012cg, SN 2014J, SN2013aa) are consistent with both Chandrasekhar-mass and sub-Chandrasekhar-mass progenitors.
Very recently, Burke et al. (2022) presented a sample of nine SNe Ia with exemplary
early-time, high-cadence, multi-wavelength follow-up from the Las Cumbres Observatory and Swift
and found that their observational results are overall consistent with
Roche-lobe-overflowing, single-degenerate progenitor systems described by companion interaction models.

H\"oflich et al.\ (2021) presented and analyzed the near-infrared (NIR) spectrum of the underluminous SN Ia 
SN 2020qxp/ASASSN-20jq, obtained with NIRES at the Keck Observatory 191 days after B-band maximum. They found good agreement between the observed lines and the synthetic profiles
computed from 3-D simulations of off-center delayed detonations in
Chandrasekhar-mass WD models.
Ashall et al. (2021) presented a multi-wavelength photometric and spectroscopic analysis of thirteen 2003fg-like SNe Ia and concluded that these
observations could be reproduced by the
CD and/or CMD scenario(s). Nevertheless, the observed signature in the late-time
nebular spectrum and light curve of SN 2006gy presented by Jerkstrand et al. (2020) supports the CMD scenario.
Siebert et al. (2021) demonstrated that SN 2019yvq is one of the best examples yet supporting the conclusion that multiple progenitor
channels may be necessary to reproduce the full diversity of normal SNe Ia. 
Various observational characteristics are presented which indicate multiple possibilities (e.g., Taubenberger 2017). In addition we note that that the properties of observed WD populations in recent sky surveys do not support many theoretical predictions in various progenitor scenarios (e.g., Rebassa-Mansergas et al. 2013, 2021; Kruckow et al. 2021; Kennea et al. 2021; Bauer et al. 2021; Korol et al. 2022; Lagos et al. 2022; Hernandez et al. 2022).

The detection of electromagnetic signals in the radio and X-ray bands from the interaction between
circumstellar material (CSM) and SN ejecta provides crucial clues for constraining progenitor models:
the SD, CD and CMD scenarios with non-degenerate companions are more likely to produce a
hydrogen-rich or helium-rich CSM. Moreover, a very recent observational study of SN 2020eyj (Kool et al. 2022) showed the helium-rich circumstellar material (CSM) interaction in SN 2020eyj, and they suggest the SD scenario with helium star donors (e.g., Ablimit 2022) might be the progenitor of SN 2020eyj. Interestingly, the expected CSM from CMD scenario (especially with helium donors, see Ablimit [2011] for more details) may also produce the properties of SN 2020eyj.
Observations of SN 2002ic, SN 2005gj, SN 2006X, SN
2008J, PTF 11kx, SN
2015cp and SN 2018fhw (e.g. Hamuy et al. 2003; Aldering et al. 2006;
Patat et al. 2007; Taddia et al. 2012; Dilday et al. 2012; Graham et al. 2019;
Vallely et al. 2019; Kollmeier et al. 2019)
demonstrated the presence of a CSM, and the WD $+$ RG channel
(a variant of the SD scenario) has been suggested as a possible
origin for at least some of these (i.e. SN 2005gj, SN 2006X, SN
2008J, SN2015cp and SN 2018fhw). Moreover, a number of observed symbiotic recurrent
nova systems (i.e. RS Oph, T CrB and V407 Cyg) have been suggested to be observational counterparts of SN Ia progenitors in 
the WD $+$ RG channel (e.g.\ Hachisu et al. 1999; Sokoloski et al. 2006).

The evolutionary pathway of WD $+$ RG binaries has been studied for decades
(e.g., van den Heuvel et al. 1992; O'Brien et al. 2006; L$\ddot{\rm u}$
et al. 2009; Chomiuk et al. 2012; Lundqvist et al. 2020).
It has been pointed out that the relatively higher and unstable mass-transfer rate may easily lead to CE evolution, which may reduce the
contribution of the WD $+$ RG channel to SNe Ia (e.g. Yungelson \& Livio 1998; Han \& Podsiadlowski 2004).
There are many important unsolved physical processes, and the
wind mass-loss process from RG stars is one of the particularly uncertain processes in the symbiotic channel.
Some previous studies have suggested that it is a spherical stellar wind (e.g.\ Chomiuk et al. 2012; Lundqvist et al. 2020),
while other studies consider it an aspherical wind lost from the RG star
in the WD binary (O'Brien et al. 2006; L$\ddot{\rm u}$ et al. 2009).
Besides, the details of Roche-lobe (RL) mass transfer from a RG donor is still poorly understood (e.g. Pastetter \& Ritter 1989; Chen et al. 2010),
and these main physical processes need further investigation.

Magnetism may play a crucial role in the accretion and nuclear
burning processes on the WD (see Mukhopadhyay \& Bhattacharya [2022] for a recent review on magnetized compact stars). The magnetic field strength of the WD, the mass-transfer rate, the masses of the donor and the WD are the most relevant parameters for studying the effects of magnetism in WD binary evolution (e.g., Livio 1983; Ablimit
\& Maeda 2019a,b; Gupta et al. 2020; Hogg et al. 2021; Walters et al. 2021; Ablimit 2022).
Magnetized WDs have been detected in symbiotic binaries and supersoft X-ray
sources (Kahabka 1995; Sokoloski \& Bildsten 1999; Osborne
et al. 2001).
Ablimit \& Maeda (2019a) find that highly magnetized WDs, accreting H-rich material, can lead to different outcomes
in detailed WD $+$ MS binary evolution calculations for SNe Ia (see also Ablimit \& Maeda 2019b), while Ablimit (2022)
demonstrated that the contribution of the WD $+$ He star channel to SNe Ia is  moderately influenced by the magnetic field of the WD.
This suggests that the role of magnetism in WD binary evolution also needs
to be investigated for the WD $+$ RG channel.

In this work, we investigate the evolution of WD $+$ RG binaries  as a potential channel for SN Ia progenitors by considering
stable mass transfer via RL overflow (RLOF; i.e.\ avoid CE evolution), stellar-wind mass loss, non-magnetic and magnetic
WDs with the binary version of the 1D stellar-evolution code \textsc{mesa} (Modules for Experiments in Stellar Astrophysics).
The WD $+$ RG channel for SNe Ia is also commonly referred to as the symbiotic channel. In \S 2, we provide the main description of the binary physical processes and parameters in the detailed \textsc{mesa} simulations in the symbiotic channel. Results and discussions are presented in \S 3, followed by the main conclusions in \S 4.


\section{\textsc{mesa} binary stellar evolution simulations}

We simulate the detailed evolution of WD $+$ RG  binaries
using the star and binary packages of version 15140
of the \textsc{mesa} code (Paxton et al. 2011, 2015, 2019).
In the first step of our calculations, we use the star package to make a RG star module based on
a typical Pop I composition with hydrogen mass fraction $X = 0.70$, He mass fraction
$Y = 0.28$ and metallicity $Z = 0.02$. We set $initial\_zfracs = 6$ and $kappa\_file\_prefix =$ `a09' to
call the opacity tables which are built using the more recent available solar composition,
and the Henyey theory of convection with $mixing\_length\_alpha=1.8$ is used in the code.
To generate a RG star model, we start the evolution with a pre-main sequence model and continue the evolution until the central helium mass fraction is $\geq 0.98$
(at this stage it has a pure helium core and a RG radius). We construct RG star models with masses ($M_{\rm RG}$) in the range from 0.5 to 2.0\,$M_\odot$ with mass intervals of 0.1\,$M_\odot$ (see also Ablimit 2023).

With the binary package of \textsc{mesa}, we then simulate the evolution of CO WD $+$ RG binaries with initial orbital periods ($P_{\rm orb,i}$) in the range of $0.5- 2000$\,d, using the RG donor models discussed above.
For the accreting WDs, treated as point masses in the code,
two initial masses, 1.2 and 1.0\,$M_\odot$, are adopted.
All relevant mechanisms for angular-momentum evolution from the system (including magnetic braking, gravitational-wave radiation, and mass loss) during the binary evolution are taken
into account (see Paxton et al.\ 2015). If matter is lost from the system, we assume that it carries the angular momentum of either the RG or the WD, whichever is appropiate. The most important physical parameters that determine the evolution of the binaries are
the initial orbital periods, initial masses of the donor and the accretor, and the mass-transfer process. Because giant stars have low surface
gravities and extended atmospheres,  it is not so straightforward to model the RLOF mass-transfer process (see the discussions in Pastetter \& Ritter 1989; Chen et al. 2010). Here, we do not consider the complications of extended atmospheres and use the general technique developed by Kolb and Ritter (1990) to compute mass transfer,
which should yield acceptable results for stars at different evolutionary stages,
\begin{equation}
{\dot{M}_{\rm RL}} =
-\dot{M}_{0}-2{\pi}F(q_2)\frac{R^3_{\rm RL}}{GM_{\rm RG}}
\times{\int^{P_{\rm RL}}_{P_{\rm ph}} {{\Gamma _1}^{1/2}{\left(\frac{2}{\Gamma_1 +1}\right)}^{\left(\frac{\Gamma_1 +1}{2\Gamma_1 -2}\right)}
{\left(\frac{\kappa_{\rm B} T}{m_{\rm p} \mu}\right)}}\,\mathrm{d}P} ,
\end{equation}
where $\dot{M}_{\rm RL}$ is the RLOF mass-transfer rate, $\Gamma _1$ is the first adiabatic exponent,
$P_{\rm ph}$ and $P_{\rm RL}$ are the pressures at the photosphere and at the radius
when the radius of the donor is equal to its RL radius, respectively.
$T$ is the temperature of the donor,  $\kappa_{\rm B}$ is the Boltzmann constant,
and $\mu_{\rm ph}$ is the mean molecular weight.
The effective RL radius (Eggleton 1983) of the RG donor star ($R_{\rm RL}$) can be calculated as,
\begin{equation}
 R_{\rm RL} =\left(\frac{0.49q^{2/3}}{0.6q^{2/3} + {\ln} (1+q^{1/3})}\right)\,a,
\end{equation}
where $q = M_{\rm RG}/M_{\rm WD}$ and $a$ is the orbital separation.
$\dot{M}_{0}$ is
\begin{equation}
\dot{M}_{0} = \frac{2\pi}{\rm exp(1/2)}F(q_2)\frac{R^3_{\rm RL, d}}{GM_d}{\left(\frac{\kappa_{\rm B} T_{\rm eff}}{m_{\rm p} \mu_{\rm ph}}\right)^{3/2}}\rho_{\rm ph},
\end{equation}
where $m_{\rm p}$ is the proton mass, and $T_{\rm eff}$ is the effective temperature of
the donor. $\mu_{\rm ph}$ and $\rho_{\rm ph}$ are the mean molecular weight and density at its photosphere.
The fitting function $F({q_2})$ ($q_2 = M_{\rm accretor}/M_{\rm donor}$) is
\begin{equation}
\begin{array}{ll}
F(q_2) =
1.23 + 0.5\,{{\log}(q_2)}, & \qquad \textrm{for $0.5\lesssim q_2 \lesssim 10$},
\end{array}
\end{equation}
the same as in the \textsc{mesa} code. Other physical assumptions are the
same as in the instrumental \textsc{mesa} papers (e.g., Paxton et al. 2015).
There are different options for configuring mass loss for RG and AGB stars in \textsc{mesa};
we used the following wind options for the RG branch (RGB) stellar-wind mass loss,

\begin{equation}
\begin{array}{ll}
\rm{cool\_wind\_RGB\_scheme = 'Reimers'},\\
\rm{cool\_wind\_AGB\_scheme = 'Blocker'},\\
\rm{RGB\_to\_AGB\_wind\_switch = 1d-4},\\
\rm{Blocker\_scaling\_factor = 0.0003d0.}
\end{array}
\end{equation}
It is worth noting that the wind mass-loss rate before RLOF is less than
$3\times10^{-8}\,\rm{M_\odot\,yr^{-1}}$ most of the time (furthermore, only a small fraction of the mass lost
from the spherically (isotropic) stellar wind moves toward the WD); thus the RLOF mass-transfer
rate dominates for the mass accretion during the WD binary evolution.
The RLOF mass-transfer rate (${\dot{M}_{\rm RL}}$) plays the decisive role in determining 
the nature of hydrogen and helium burning on the WD and the stability of burning affects its mass
retention efficiency and how the WD grows in mass. The mass growth rate of the WD is usually written as
\begin{equation}
\dot{M}_{\rm{WD}} = \eta_{\rm H} \eta_{\rm He} {\dot{M}_{\rm RL}},
\end{equation}
where we adopt the prescription of Hillman et al. (2015, 2016) for the efficiency of
hydrogen burning ($\eta_{\rm H}$) and the methods of
Kato \& Hachisu (2004) for the mass accumulation efficiency of
helium ($\eta_{\rm He}$).
The different episodes of these burning efficiencies strongly affect
the results of the mass-accretion phase. The adopted prescriptions here are widely used, but
different models for carbon burning would somewhat affect the growth of the WD  (Brooks et al. 2017).

The stream/confined accretion and related emission on a magnetized WD has to be treated differently compared to spherically symmetric accretion onto a non-magnetic WD (e.g., Fabian et al. 1977; Livio 1983; King \& Shaviv 1984; Hameury et al. 1986; King 1993; Wickramasinghe \& Ferrario 2000; Wickramasinghe 2014;
Ferrario et al. 2015; Mukhopadhyay et al. 2017; Ablimit 2019, 2022).
For a sufficiently magnetized WD, the mass flow onto the WD, once RLOF has started, will be magnetically channeled and not through an accretion disk connected to the WD (e.g., Cropper 1990; Frank et al. 2002; see the schematic figure of Ablimit \& Maeda (2019a) and Ablimit (2019)). The strong magnetic field of the WD controls the motion of the accreting matter near the WD; as the WD's magnetic
pressure increases more rapidly than the ram pressure of the accreting material as it approaches the WD's surface, there will be a radius, the magnetospheric radius, at which the magnetic pressure is equal to the ram pressure (Frank et al. 2002). Below this radius, matter will flow along magnetic field lines  and fall onto the magnetic poles of the WD through an accretion column.
The minimum physical condition for magnetically confined accretion is that the magnetic field strength ($B$) satisfies (Livio 1983),
\begin{equation}
{B} \geq 9.3\times10^{7} \left(\frac{{R_{\rm WD}}}{5\times10^8\,\rm cm}\right)\,\left(\frac{{P_{\rm b}}}{5\times10^{19}\,\rm{dyne\,cm^{-2}}}\right)^{7/10}\,
\left(\frac{{M_{\rm WD}}}{ M_\odot}\right)^{-1/2}\left(\frac{\dot{M}}{10^{-10}\,{M_\odot\,\rm{yr}^{-1}}}\right)^{-1/2}\, \rm{G},
\end{equation}
 where $\dot{M}$ is the RLOF mass transfer rate ($\dot{M}_{\rm RL}$ in this work ).
The pressure at the base of the accreted matter ($P_{\rm b}$) is related to the properties of the WD, the size of the polar cap regions, and the accreted mass, and is taken as ${P_{\rm b}} = 5\times10^{19}\,\rm{dyne\,cm^{-2}}$ (following Livio (1983)).
We also adopt the mass ($M_{\rm WD}$) -- radius ($\rm{R_{WD}}$) relation of Nauenberg (1972) for the WD. 

For WDs with no magnetic field or an intermediate-strength magnetic field, the
mass-transfer rate has to be $\ga 5\times10^{-8}\,M_{\odot}\,{\rm yr^{-1}}$  to avoid nova outbursts (e.g., Hillman et al. 2015, 2016).
Once the magnetic field strength of the WD meets the condition for magnetic confinement,  nova
outburst can be suppressed at a much lower mass-transfer rate.
In this work, we adopt the approach of Ablimit \& Maeda (2019a) and Ablimit (2019, 2022) for simulating the accretion phase of the WD,
and consider WDs with no magnetic fields and WDs with intermediate and high magnetic field strengths ($B$) and the effects of the magnetic fields on the binary evolution.
We take a sufficiently magnetized WD with a fixed initial magnetic field strength of
$2.5\times10^7$\,G to realize magnetic confinement (cf.\ Livio 1983).
For the purposes of deciding whether nova outbursts occur and to calculate the accretion efficiencies in Eq.~6, we define an equivalent, isotropic polar mass-transfer rate ($\dot{M}_{\rm{p}}$) in the case of magnetic confinement as (see also Ablimit \& Maeda 2019a),
\begin{equation}
\dot{M}_{\rm{p}} =  \frac{S}{\Delta {S}} {\dot{M}_{\rm RL}},
\end{equation}
where the ratio of the surface area of the WD and the two polar regions of the WD (on which material is accreted) is ${S}/{\Delta {S}}={{2 R_{\rm m}}}/({\rm{R_{WD}} \rm{cos^2{\theta}})}$, where we take $\theta = 0$ (the angle between the rotation axis and the magnetic field axis; see Ablimit (2022) for more information) for the simulations in this work.
${R_{\rm m}}$ is the magnetospheric radius which is related with Alfven radius ${R_{\rm A}}$ (Lamb et al.\ 1973; Norton \& Watson 1989; Frank et al.\ 2002),
\begin{equation}
R_{\rm m}= \phi {R_{\rm A}} = \phi\, 2.7\times10^{10} {M^{-1/7}_{\rm WD}} {\dot{M}^{-2/7}_{\rm acc}} {{\mu}^{4/7}}\,\rm{cm}, 
\end{equation}
where $\phi$ is a parameter ($\leq 1$) that takes into account the departure from the spherically symmetric case,  ${\mu}= {B} {R^3_{\rm WD}}$ is the magnetic moment of the WD 
in units of $10^{33}$ G $\rm{cm}^3$, $\dot{M}_{\rm acc}$ is the mass-accretion rate in units of $10^{16}$ g $\rm{s}^{-1}$ and $M_{\rm WD}$ is the WD mass in solar units.


\section{Results and discussion}

The Hertzsprung-Russell (HR) diagram in Figure 1 shows that the RG stars modeled with the \textsc{mesa} code
in this work fit with our current understanding of stellar evolution (note that the figure only shows theoretical evolution tracks without observational constraints).
Figure 2 shows the outcomes of the binary evolution sequences in the initial RG donor mass -- initial orbital period plane, where the blue solid lines enclose the 
regions that produce SNe Ia (i.e. regions where the WDs grow in mass and reach the Chandrasekhar mass limit, taken as $M_{\rm Ch}= 1.38 M_\odot$)), for different magnetic field strengths.
The upper panels of Figure 2 are for the initial parameter
space of CO WD + RG star binaries with $M_{\rm WD,i}=1.2$ and $1.0\,M_\odot$
with no magnetic fields or intermediate magnetic field strength, for which no magnetic confinement occurs. The ranges of initial donor mass and orbital
period for the case without magnetic confinement with $M_{\rm WD,i}=1.2$ and $1.0\,M_\odot$ are
$1.1-1.8\,M_\odot$ and $10-750$\,d, and $1.3-1.7\,M_\odot$ and $35-400$\,d, respectively.
For the highly magnetized WDs with $M_{\rm WD,i}=1.2$ and $1.0\,M_\odot$
 (with magnetic confinement; lower panels of Figure 2),
they are $0.7-1.8\,M_\odot$ and $5-120$\,d, and $1.2-1.7\,M_\odot$ and $20-28$\,d, respectively.
Compared to previous similar studies
(e.g., Li \& van den Heuvel 1997; L$\ddot{\rm u}$ et al. 2009; Wang \& Han 2010; Liu et al. 2019),
the ranges from both models (those with and without magnetic
confinement) are different for the following reasons:
1. The treatment of mass transfer varies for the different stellar evolution codes. 
Here, we use the \textsc{mesa} code with the Kolb scheme for the RLOF mass transfer.
2. Both wind mass loss and RLOF mass transfer are considered at the same time
in our simulations. This is important as  RG stars can experience substantial wind mass loss,
which tends to widen the orbits prior to the beginning of RLOF.
Some RG stars (near the upper mass range) can lose up to  $\sim 32$\% of their mass through their stellar winds, which
can increase the orbital period by $\sim 21$\% prior to RLOF. This may help to stabilize the RLOF mass transfer in some binaries.
We take both wind mass loss and RLOF mass transfer
into account while previous studies only considered one of them.
Compared to the case without magnetic confinement, the initial donor mass can be lower
for the high magnetic-field case, because magnetic confinement leads to a higher $\dot{M}_{\rm p}$, which in turn allows matter to burn more stably and increases
$\dot{M}_{\rm WD}$ even in lower-mass donor stars.
The ranges of initial orbital periods that produce SNe Ia shrink because 
the mass loss from the system is generally higher for the higher values of $\dot{M}_{\rm p}$ (the WD eject some of the transferred mass when $\dot{M}_{\rm p}> 10^{-6}\,M_\odot\,{\rm yr^{-1}}$).

Figure 3 shows the detailed binary evolution of
one WD + RG system without and with magnetic confinement.
Without magnetic confinement (left panels of Figure 3), the RG 
with an initial mass of 0.8\,$M_\odot$ cannot let the WD ($M_{\rm WD,i}=1.2\,M_\odot$) grow in mass
to the Chandrasekhar mass limit as
the low mass-transfer rate leads to nova outbursts with no significant accretion.
With the effect of the WD's strong magnetic field (right panels of Figure 3),
the transferred matter can be confined to the polar cap regions, and the
higher polar mass-accretion rate (red line) alters the mass-accretion phase of the binary.
The main difference of these two cases with
the same RLOF mass transfer from the donor is that the hydrogen burning efficiency (mass retention efficieny) on the WD due to the
magnetic confinement is higher than that in the spherical accretion case.
As the RG star loses its mass, the WD gains mass smoothly due to the magnetic
confinement, and this mass accretion leads to a contraction of the orbit.
In contrast, the WD's mass remains constant (i.e., experiences no significant accretion) without magnetic confinement, and the transferred mass will be lost from the binary,
and this mass loss, including the RG stellar wind mass loss, will take angular momentum with it.
Thus, the orbital period evolves from 10 to 75
days in the non-magnetic case.
In the magnetic confinement case, the orbital period evolves in a much slower way
because the system loses only a small amount of mass through the RG stellar
wind (the WD accretes the mass transferred through RLOF from the RG).
Because of the different mass accretion,
the radius of the RG star expands more than in the case  of a RG without magnetic confinement.

In Figure 4, we show another example for the evolution of a WD $+$ RG  binary,
where the WD ($M_{\rm WD,i}=1.2\,M_\odot$) can reach $M_{\rm Ch}$ with
and without magnetic confinement,
and the mass transfer in both cases proceeds on a
thermal timescale. The main difference between the evolution of the two cases is that
the accretion rate per unit area on the WD is higher with magnetic confinement.
In this binary, the wind mass-loss rate of the RG star with an initial
mass of 1.1\,$M_\odot$ is higher than
$10^{-10}\,M_\odot\,{\rm yr^{-1}}$ and can be as high as $10^{-8}\,M_\odot\,{\rm yr^{-1}}$;
it takes more angular momentum away from the binary; thus, the wind
mass loss widens the orbit significantly compared to the previous example.
With the higher-mass donor (higher
RLOF mass-transfer rate), the polar mass-transfer rate will be
higher, and the outcomes will accordingly be different.
In the magnetic confinement case (right panels), some of the transferred
 mass via RLOF would be lost from the binary and take away angular momentum;
the mass loss rate will be higher if $\dot{M}_{\rm p}$ (red line) is higher (when it exceeds $10^{-6}\,M_\odot\,{\rm yr^{-1}}$), and some of the transferred matter can not burn stably on the WD and escape from the WD. Thus, the RG donor
loses more mass in order to allow the WD to grow in mass to $M_{\rm Ch}$,
and this mass loss also widens the binary orbit
more compared to the case of no magnetic confinement (left panels).

There are some
differences in other properties (i.e., radius, see Figure 4; luminosity and effective
temperature, see Figure 5) of the
giant donors in the two cases due to the different mass accretion/mass loss (or/and different initial conditions).
The giant donor in the magnetic
confinement case loses more mass and evolves further towards lower mass than the final donor
in the case without magnetic confinement. Thus, the total ranges of orbital period and
donor properties (mass, orbital velocity, luminosity, effective temperature) at the time of the SN explosion are different in the two cases.
For the case without magnetic confinement, combining the results for $M_{\rm WD,i}=1.2$ and $1.0\, M_\odot$, the ranges of orbital periods, donor mass, orbital velocity,
luminosity and effective temperature are $21.3-2521.5$\,d, $0.59-1.23$\,$M_\odot$, $13.5-57.7$\,km\,$\rm s^{-1}$, $1.94-3.62$ (in $\log (L/L_\odot)$), and $2960-4458$\,K, respectively, while for the magnetic confinement case they are $5-652$\,d, $0.41-1.21\,M_\odot$, $22.3-100.5$\,km\,$\rm s^{-1}$,
$1.08-2.83$ (in $\log(L/L_\odot)$), and $3268-4700$\,K, respectively.
In both examples, the ages of the RG stars are very long, implying that the WD + RG channel has
a long delay time between the star-formation phase and the time of the SN explosion, and hence this channel would contribute to the
old population of SNe Ia. The giant stars will survive after the SN explosion and finally evolve to become WDs.
The existence of single WDs (especially low-mass He WDs with $< 0.45$\,$M_\odot$) could be the survivors from SNe Ia
produced by this channel. Indeed, a population of low-mass, apparently single WD, presumably He WDs, has been found in the cluster NGC 6791 (Kilic et al. 2007) and in the field (Kawka et al.\ 2006; Bergeron \& Leggett 2002) and have been proposed to be SN Ia survivors (Justham et al.\ 2009).
Besides, some observed symbiotic novae appear to occur in binaries with very massive WDs and relatively low-mass giant companions; these systems are potential observational counterparts of SN Ia progenitors, e.g.,  RS Oph (e.g., Brandi et al. 2009; Mikolajewska \&
Shara 2017), T CrB (e.g., Belczynski \& Mikolajewska 1998), V745 Sco (e.g., Drake et al. 2016; Orlando et al. 2017).
The properties of these symbiotic nova systems can be well reproduced by the
WD $+$ RG binary evolutionary sequences with and without magnetic confinement.
For future observations, it will be very interesting and challenging to find
the surviving He WDs or stripped giant stars whose envelopes have been partially lost by the stellar wind or/and during the mass-transfer phase of in the supernova explosion, which would be less massive and hotter than comparable single stars.
In addition, detailed observations of SN Ia environments can provide further clues, and the CSM properties from SNe Ia
observations may constrain this model as the wind mass loss may create a CSM-like
environment around the system in this symbiotic channel (but see also Moriya et al. 2013).
In addition, some observational clues such as continued observations of late-time light curves of nearby SNe Ia, new high-cadence survey capacities
in the short term (Ivezi$\acute{c}$ et al.\ 2019) and future direct observations
via gravitational waves (Korol et al.\ 2018) will provide crucial information on the nature of the SN Ia progenitors. Additionally, the evolution of oxygen-neon-magnesium composition WD -- RG binaries via accretion-induced collapse provides a promising channel to form peculiar neutron star X-ray binaries (Ablimit 2023).

\section{Conclusion}

As already discussed in the Introduction, both observational and theoretical studies to date suggest that a number of progenitor models may  contribute to the SN Ia population.
In this study, we employed the \textsc{mesa} stellar evolution code to simulate a large grid of WD $+$ RG binaries as potential SN Ia progenitors, 
considering WDs without magnetic field, with intermediate, and high magnetic-field strengths in these binaries.
In the simulations, the RLOF mass-transfer rate is always higher than the wind mass-loss rate, and the mass-accretion phase is dominated by RLOF mass transfer, while the stellar wind causes mass loss from the systems (which mainly affects the orbital period by taking away angular momentum).

Compared to systems with WDs with no or intermediate magnetic-field strength (where magnetic confinement does not occur), highly magnetized WDs can confine the
transferred matter to their polar caps and can increase the burning efficiency even with a lower mass-transfer rate.
With magnetic confinement, systems with a lower-mass RG star can drive the WD to grow in mass to experience a SN explosion,
as shown in the initial parameter space of the RG mass and orbital period.
In the case without magnetic confinement, the initial parameter space (initial donor mass and initial orbital
period) derived in this study for producing SNe Ia (for $M_{\rm WD,i}=1.2$ and $1.0\,M_\odot$) are $1.1-1.8\,M_\odot$ and $10-750$\,d, and $1.3-1.7\,M_\odot$ and $35-400$\,d, respectively.
For the highly magnetized WDs (for $M_{\rm WD,i}=1.2$ and $1.0\,M_\odot$),
they are $0.7-1.8\,M_\odot$ and $5-120$\,d, and $1.2-1.7\,M_\odot$ and $20-28$\,d, respectively.
Based on the two examples for which we derived the post-SN properties,
we suggest that finding a He WD or a RG with an
envelope that has at least partially been stripped, lower-mass giant (or/and hotter) stars after
the SN or inside a SN remnant will provide a conclusive test for this symbiotic scenario.
In the future we will also consider the possibility of super-Chandrasekhar WDs and the contribution of highly magnetized WDs in binaries with MS, RG and stripped helium star companions to the class of overluminous SNe Ia (Ablimit et al. in preparation).

\software{Modules for Experiments in Stellar Astrophysics (MESA; Paxton et al. 2011, 2015, 2019)}

\section*{Acknowledgments} We thank Noam Soker for useful comments and discussions. This work is supported by NSFs. 

\section*{Data Availability} The data underlying this article
will be shared on reasonable request to the corresponding author.

%




\clearpage

\begin{figure}
\centering
\includegraphics[totalheight=4.5in,width=5.8in]{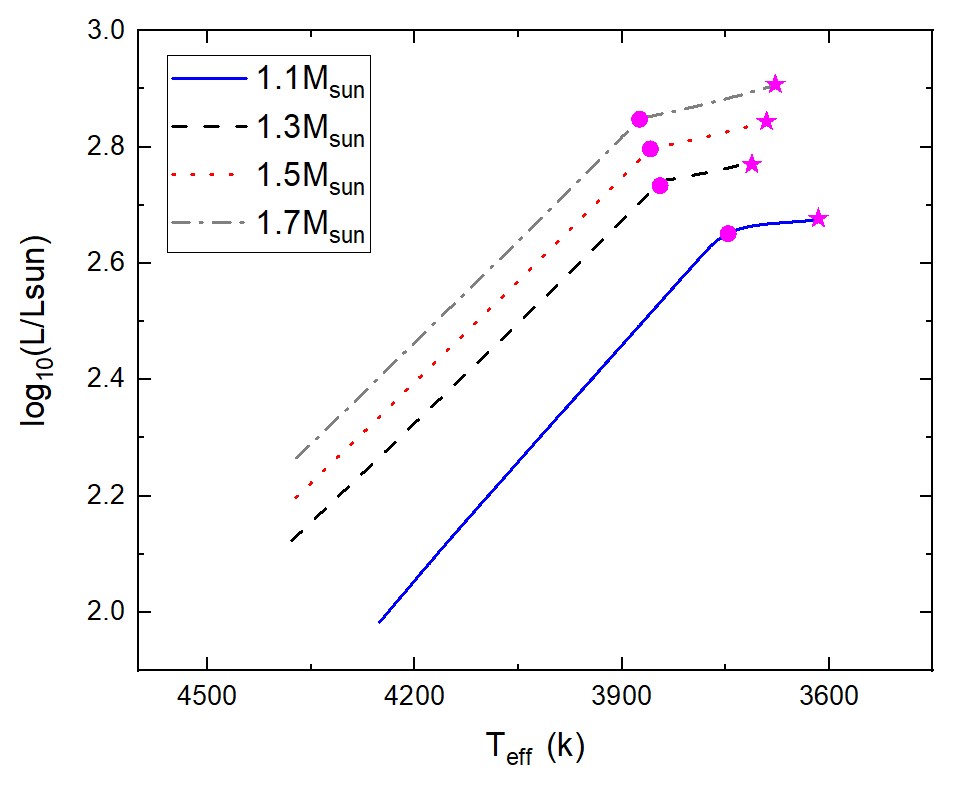}
\caption{Hertzsprung-Russell diagram for red-giant donors with different masses in non-magnetized CO WD binaries.
Red circles indicate the location where the WDs starts to accrete matter, and red stars the location when SNe Ia occur.}
\end{figure}

\clearpage

\begin{figure}
\centering
\includegraphics[totalheight=4.5in,width=5.8in]{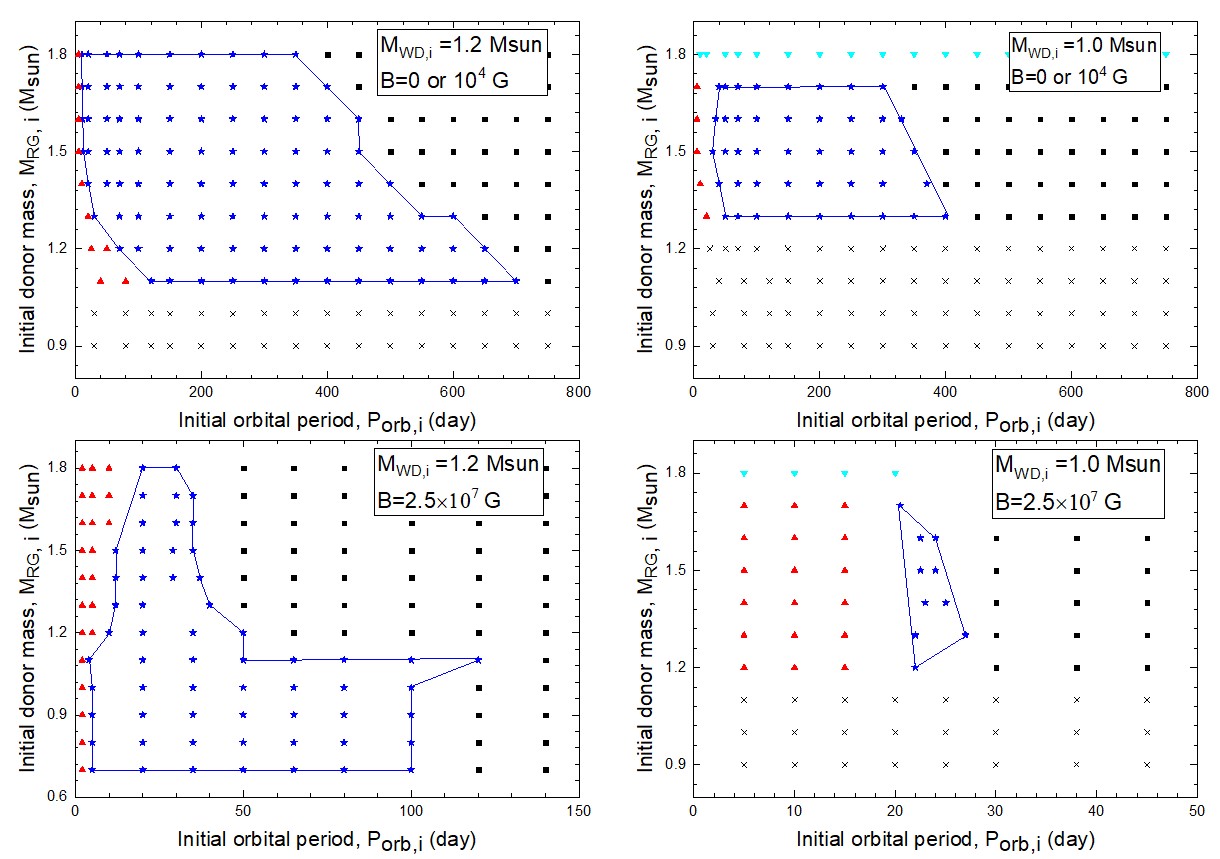}
\caption{Initial parameter space for producing SNe Ia: the initial orbital period ($P_{\rm orb,i}$)--initial red-giant donor mass ($M_{\rm RG,i}$) plane
for WD $+$ RG systems. The initial WD masses are 1.0 and 1.2\,$M_\odot$. The panels show the results
for WDs with no or intermediate-strength magnetic fields (upper panel), and high magnetic fields (lower panel), respectively.
Blue stars show the cases that produce SNe Ia. The solid blue lines show the parameter regions within which SNe Ia are produced. 
The cyan triangles in the upper regions
in the figure indicate the occurrence of a common
envelope, the black crosses in the lower regions show were nova outbursts occur.
The red triangles and black squares on the sides show regions with inefficient mass transfer.
}
\end{figure}

\clearpage

\begin{figure}
\centering
\includegraphics[totalheight=6.5in,width=5.8in]{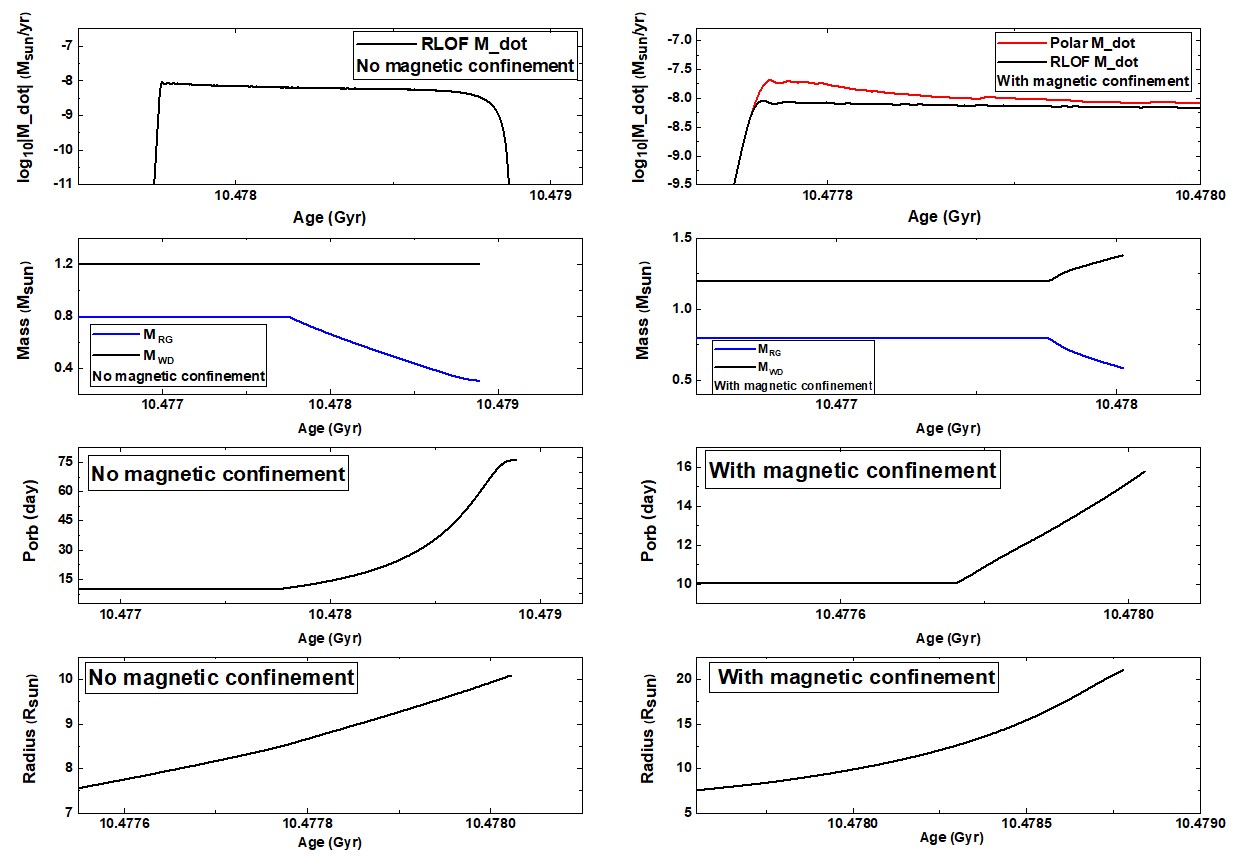}
\caption{Detailed evolution of CO WD $+$ RG binaries as a function of time using the \textsc{mesa}
code. The left panels show the case of a non-magnetic WD and the right panels the case of a magnetized WD (with $B=2.5\times10^7$\,G). All other initial parameters are the same: the
initial masses of the WDs and donor stars are 1.2\,$M_\odot$ and 0.8\,$M_\odot$, respectively,  with an initial orbital period of 10\,d.
The evolution of mass transfer, WD/donor mass, orbital period, and the radius of the donors are shown.
Wind mass loss is very low ($<5\times10^{-13}\,M_\odot/{\rm yr}$) and is not shown for clarity. RLOF M\_dot and Polar M\_dot in the figures are $\dot{M}_{\rm RL}$ and $\dot{M}_{\rm{p}}$ in the text, respectively.} 
\end{figure}

\clearpage

\begin{figure}
\centering
\includegraphics[totalheight=6.5in,width=5.8in]{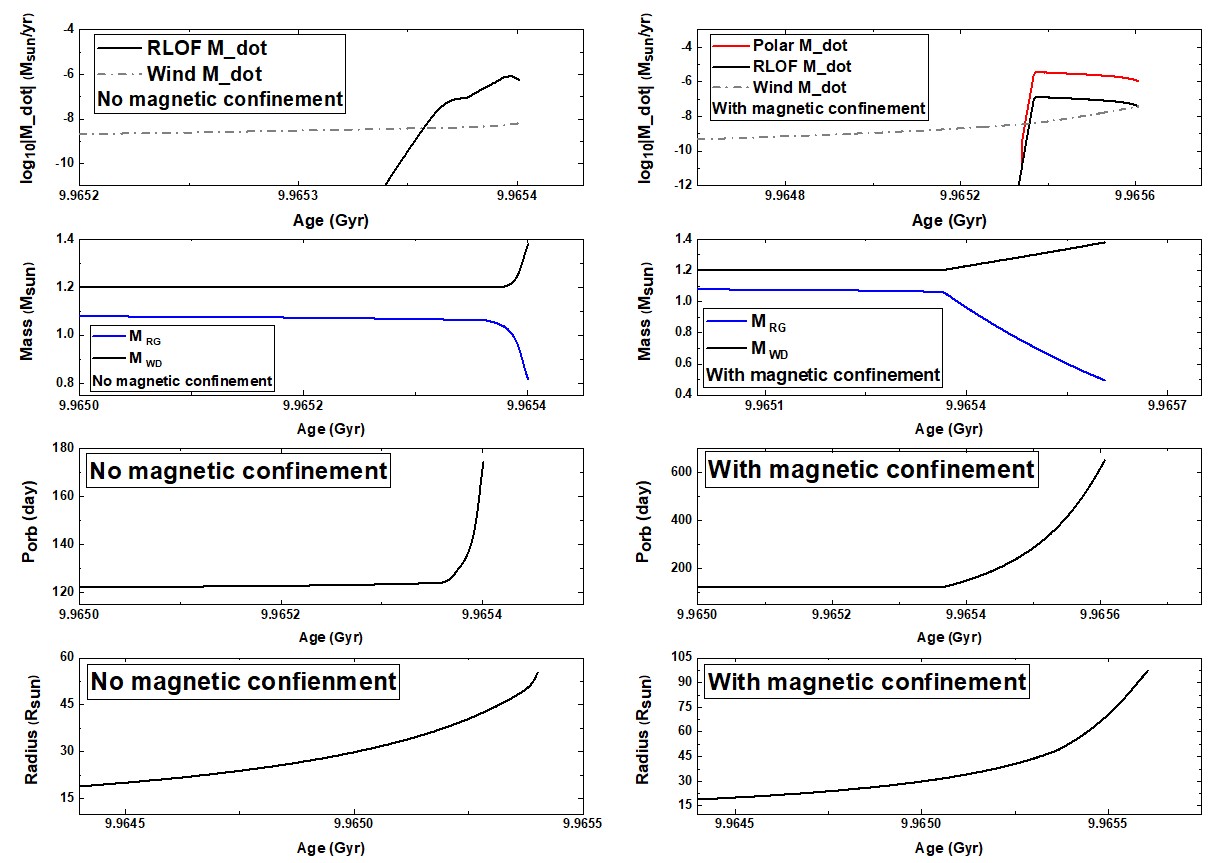}
\caption{Another example for the detailed evolution of CO WD $+$ RG binaries using the \textsc{mesa} code: evolution of mass transfer, wind mass loss,
WD/donor mass, orbital period, and radius of the donors  are shown as a function of time. The right and left panels are for a non-magnetized and highly magnetized WD ($B=2.5\times10^7$\,G) with magnetic confinement, respectively. The
initial masses of the WD and RG donor are 1.2\,$M_\odot$ and 1.1\,$M_\odot$, and the initial orbital period is 120 days for both cases. }
\end{figure}

\clearpage

\begin{figure}
\centering
\includegraphics[totalheight=4.5in,width=5.8in]{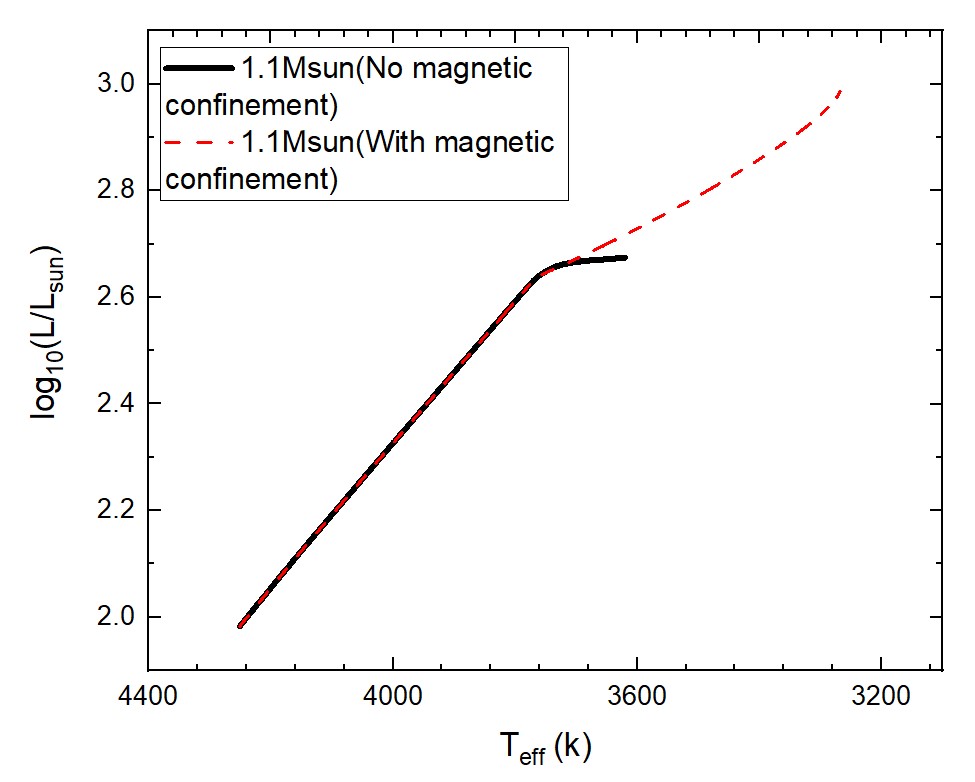}
\caption{Example of the evolution of a RG star in the Hertzsprung-Russell diagram for the CO WD + RG binaries without and with magnetic confinement.}
\end{figure}


\clearpage

\section*{REFERENCES}

Ablimit, I., 2021, PASP, 133, 074201, doi:10.1088/1538-3873/ac025c

Ablimit, I., 2022, MNRAS, 509, 6061, doi:10.1093/mnras/stab3060

Ablimit, I. 2023, MNRAS, 519, 1327, doi:10.1093/mnras/stac3551

Ablimit, I., Maeda, K. \& Li, X.-D., 2016, ApJ, 826, 53, doi:10.3847/0004-637X/826/1/53

Ablimit, I., \& Maeda, K. 2019a, ApJ, 871, 31, doi:10.3847/1538-4357/aaf722

Ablimit, I., \& Maeda, K. 2019b, ApJ, 885, 99, doi:10.3847/1538-4357/ab4814

Ablimit, I., Xu, X.-J \& Li, X.-D. 2014, ApJ, 780, 80, doi:10.1088/0004-637X/780/1/80

Aldering, G., Antilogus, P., Bailey, S., et al., 2006, ApJ, 650, 510, doi:10.1086/507020

Ashall, C., Lu, J., Hsiao, E. Y. et al., 2021, arXiv:2106.12140

Bauer, E. B.; Chandra, V. Shen, K. J. et al. 2021, ApJ, 923, L34, doi:10.3847/2041-8213/ac432d

Belczynski, K., \& Mikolajewska, J. 1998, MNRAS, 296, 77, doi:10.1046/j.1365-8711.1998.01301.x

Bergeron, P.\ \& Leggett, S. K. 2002, ApJ, 1070

Brooks, J., Schwab, J., Bildsten, L., Quataert, E., \& Paxton, B. 2017, ApJ,
843, 151, doi:10.3847/1538-4357/aa79a6

Brandi, E., Quiroga, C., Mikolajewska, J. et al., 2009, A\&A, 497, 815, doi:10.1051/0004-6361/200811417

Burke, J., Howell, A., Sand, D. J. et al., arXiv: 2207.07681

Chen, X., Podsiadlowski, Ph., Mikolajewska, J., \& Zhanwen, H., 2010, AIP Conf. Proc., Vol.1314, p. 59

Claeys, J. S. W., Pols, O. R., Izzard, R. G., Vink, J. \& Verbunt, F. W. M. 2014, A\&A, 563, A83, doi:10.1051/0004-6361/201322714

Cropper, M., 1990, Space Sci. Rev., 54, 195, doi:10.1007/BF00177799

Dilday, B., Howell, D., Cenko, S., 2012, Sci, 337, 924, doi:10.1126/science.1219164 

Dimitriadis, G. et al., 2019, ApJL, 870, L1, doi:10.3847/2041-8213/aaedb0 

Drake, J. J., Delgado, L., Laming, J. M., et al., 2016, ApJ, 825,
95, doi:10.3847/0004-637X/825/2/95

Di Stefano, R. \& Kilic, M. 2012, 759, 56, doi:10.1088/0004-637X/759/1/56

Eggleton, P. P. 1983, A\&A, 268, 368, doi:10.1086/160960

Fabian, A. C., Pringle, J. E., Rees, M. J., Whelan, J. A. J., 1977, MNRAS, 179,
9, doi:10.1093/mnras/179.1.9P 

Fausnaugh, M. M. et al., 2019, Preprint at https://arxiv.org/abs/1904.02171

Ferrario, L., de Martino, D., Gaensicke, B. T., 2015, Space Sci. Rev., 191, 111F, doi:10.1007/s11214-015-0152-0 

Frank, J., King, A., Raine, D. J., 2002, in Frank J., King A., Raine D., eds,
Accretion Power in Astrophysics. Cambridge Univ. Press, Cambridge, p.
398

Fryer, C. L., Ruiter, A. J., Belczynski, K. et al. 2010, ApJ, 725, 296, doi:10.1088/0004-637X/725/1/296 

Graham, M., Harris, C., Nugent, P., et al., 2019, ApJ, 871, 62, doi:10.3847/1538-4357/aaf41e

Gupta, A., Mukhopadhyay, B. \& Tout, C. A. 2020, MNRAS, 496, 894, doi:10.1093/mnras/staa1575

Hachisu, I., Kato, M., Nomoto, K., 1999, ApJ, 522, 487, doi:10.1086/307608

Hameury, J.-M., King, A. R., Lasota, J.-P., 1986, MNRAS, 218, 695, doi:10.1093/mnras/218.4.695

Hamuy, M., et al., 2003, Nat, 424, 651, doi:10.1038/nature01854

Han, Z. \& Podsiadlowski, Ph. 2004, MNRAS, 350, 1301, doi:10.1111/j.1365-2966.2004.07713.x

Hernandez, M. S., Schreiber, M. R., Parsons, S. G. et al. 2022, MNRAS, 517, 2867, doi:10.1093/mnras/stac2837 

H\"oflich, P., Ashall, C., Bose, S. et al., 2021, arXiv:2109.03359

Hogg, M. A., Cutter, R., Wynn, G. A. 2021, MNRAS, 500, 2986, doi:10.1093/mnras/staa3316

Hillman, Y., Prialnik, D., Kovetz, A., \& Shara, M. M. 2015, MNRAS, 446, 1924, doi:10.1093/mnras/stu2235

Hillman, Y., Prialnik, D., Kovetz, A., \& Shara, M. M. 2016, ApJ, 819, 168, doi:10.3847/0004-637X/819/2/168

Hristov, B., Collins, D. C., Hoeflich, P. et al. 2018, ApJ, 858, 13

Ivezi$\acute{c}$, $\breve{Z}$., Kahn, S. M., Tyson, J. A., et al., 2019, ApJ, 873, 111, doi:10.3847/1538-4357/ab042c 

Jerkstrand, A., Maeda, K., \& Kawabata, K. S. 2020, Science, 367, 415, doi:10.1126/science.aaw1469

Justham, S., Wolf, C., Podsiadlowski, Ph., \& Han, Z. 2009, A\&A, 493, 1081

Jha, S. W., Maguire, K., \& Sullivan, M. 2019, Nature Astronomy, 3, 706, doi:10.1038/s41550-019-0858-0

Kahabka, P. 1995, ASP Conference Series, Vol. 85

Kasen, D., Roepke, F. K. \& Woosley, S. E. 2009, Nat., 460, 869, doi:10.1038/nature08256

Kawka, A., Vennes, S., Oswalt, T. D., Smith, J. A., \& Silvestri, N. M. 2006, ApJ,
643, L123

Kashi, A., Soker, N., 2011, MNRAS, 417, 1466, doi:10.1111/j.1365-2966.2011.19361.x

Kennea, J. A., Coe, M. J., Evans, P. A. et al. 2021, MNRAS, 508, 781, doi:10.1093/mnras/stab2632

Kilic, M., Stanek, K. Z., \& Pinsonneault, M. H. 2007, ApJ,
671, 761, doi:10.1086/522228 

King, A. R., 1993, MNRAS, 261, 144, doi:10.1093/mnras/261.1.144

King A. R., Shaviv G., 1984, MNRAS, 211, 883, doi:10.1093/mnras/211.4.883

Khokhlov, A. M. 1991, A\&A, 245, L25, doi:

Kolb, U., \& Ritter, H. 1990, A\&A, 236, 385

Kollmeier, J., Chen, P., Dong, S., et al., 2019, MNRAS, 486, 3041, doi:10.1093/mnras/stz953

Kool1, E. C.,  Johansson, J., Sollerman, J. et al. 2022, arXiv:2210.07725

Korol, V., Koop, O., \& Rossi, E. M. 2018, ApJL, 866, L20, doi:10.3847/2041-8213/aae587

Korol, V. Belokurov, V. \& Toonen, S. 2022, MNRAS, 515, 1228, doi:10.1093/mnras/stac1686

Kruckow, M. U., Neunteufel, P. G., Di Stefano, R. et al. 2021, ApJ, 920, 86, doi:
10.3847/1538-4357/ac13ac

Kushnir, D., Katz, B., Dong, S., Livne, E., \& Fernandez, R. 2013, ApJ, 778, L37

Lach, F., Callan, F. P., Bubeck, D. et al., 2021, arXiv:2109.02926

Lamb, F. K., Pethick, C. J. \& Pines, D., 1973, ApJ, 184, 271, doi:10.1086/152325

Langer, N., Deutschmann, A., Wellstein, S. \& H\"oflich, P. 2000, A\&A, 362, 1046

Li, X. D., van den Heuvel, E. P. J., 1997, A\&A, 322, L9

Liu, D., Wang, B., Ge, H., Chen, X., Han Z., 2019, A\&A, 622, A35

Livio, M. 1983, A\&A, 121, L7

Livio, M. \& Mazzali, P. 2018, Physics Reports, 736, 1, doi:10.1016/j.physrep.2018.02.002

Liu, Z. W., Roepke, F. K., Zeng, Y. \& Heger, A. 2021, A\&A, 654, 103, doi:10.1051/0004-6361/202141518 

Lagos, F., Schreiber, M. R., Parsons, S. G. et al. 2022, MNRAS, 512, 2625, doi:10.1093/mnras/stac673 

L$\ddot{\rm u}$, G., Zhu, C., Wang, Z., \& Wang, N. 2009, MNRAS, 396, 1086

Maeda, K., \& Terada, Y. 2016, IJMPD, 25, 1630024, doi:10.1142/S021827181630024X

Magee, M. R., Sim, S. A., Kotak, R. \& Kerzendorf, W. E. 2018, A\&A, 614, A115, doi:10.1051/0004-6361/201832675

Maoz, D., Mannucci, F., Nelemans, G., 2014, ARA\&A, 52, 107, doi:10.1146/annurev-astro-082812-141031

Marsh, T. R., Dhillon, V. S., \& Duck, S. R. 1995, MNRAS,
275, 828, doi:10.1093/mnras/275.3.828

Meng, X., \& Podsiadlowski, Ph. 2017, MNRAS, 469, 4763

Miko\l{}ajewska, J., \& Shara, M. M. 2017, ApJ, 847, 99, doi:10.3847/1538-4357/aa87b6

Moriya, T., Maeda, K., Taddia, F., et al., 2013, MNRAS, 435, 1520, doi:10.1093/mnras/stt1392

Mukhopadhyay, B., Rao, A. R. \& Bhatia, T. S. 2017, MNRAS, 472, 3564, doi:10.1093/mnras/stx2119

Mukhopadhyay, B. \& Bhattacharya, M. 2022, Particles 5, 493, doi:10.3390/particles5040037 

Nauenberg, M. 1972, ApJ, 175, 417, doi:10.1086/151568

Nelemans,  G.,  Toonen,  S.,  \&  Bours,  M.  2013,  in
Binary  Paths  to  Type  Ia Supernovae Explosions, IAU Symp., 281, 225

Norton, A.J., Watson, M.G. 1989, MNRAS, 237, 715, doi:10.1093/mnras/237.3.715

O'Brien, T. J. et al., 2006, Nat, 442, 279, doi:10.1038/nature04949

Orlando, S., Drake, J. J., \& Miceli, M. 2017, MNRAS, 464,
5003, doi:10.1093/mnras/stw2718 

Osborne,  J. P., Borozdin, K. N., Trudolyubov, S. P. et al., 2001, A\&A, 378, 800, doi:10.1051/0004-6361:20011228

Pakmor, R., Kromer, M., Taubenberger, S. \& Springel, V. 2013, ApJ, 770, L8, 
doi:10.1088/2041-8205/770/1/L8

Pastetter, L., \& Ritter, H. 1989, A\&A, 214, 186

Patat, E., et al., 2007, Sci, 317, 924

Paxton, B., Bildsten, L., Dotter, A., Herwig, F., Lesaffre, P. \& Timmes, F. 2011, ApJS, 912, 3, doi:10.1088/0067-0049/192/1/3

Paxton, B., Marchant, P., Schwab, J. et al., 2015, ApJS, 220, 15, doi:10.1088/0067-0049/220/1/15

Paxton, B., Smolec, R., Schwab, J. et al., 2019, ApJS, 243, 10, doi:10.3847/1538-4365/ab2241

Perets, H. B., Zenati, Y., Toonen, S., Bobrick, A. 2019, eprint arXiv:1910.07532

Piro, A. L. \& Nakar, E. 2013, ApJ, 769, 67, doi:10.1088/0004-637X/769/1/67

Podsiadlowski, Ph., Mazzali, P., Lesaffre, P. et al.,  2008, NewAR, 52, 381, doi:10.1016/j.newar.2008.06.020 

Polin, A., Nugent, P. \& Kasen, D. 2019,  ApJ, 873, 84, doi:10.3847/1538-4357/aafb6a

Raskin, C., Timmes, F. X., Scannapieco, E., Diehl, S., \& Fryer,C. 2009, MNRAS, 399, L156

Rebassa-Mansergas, A., Agurto-Gangas, C., Schreiber, M. R. et al. 2013, MNRAS, 433, 3398, 
doi:10.1093/mnras/stt974

Rebassa-Mansergas, A., Solano, E., Jimenez-Esteban, F. M. et al. 2021, MNRAS, 506, 5201, 
doi:10.1093/mnras/stab2039

Roepke, F. K. \& Niemeyer, J. C. 2007, A\&A, 464, 683, doi:10.1051/0004-6361:20066585

Roepke, F. K.; Woosley, S. E.; Hillebrandt, W. 2007, ApJ, 660, 1344, doi:10.1086/512769

Ruiter, A. J., Belczynski, K. \& Fryer, C. 2009, ApJ, 699, 2026, doi:10.1088/0004-637X/699/2/2026

Sato, Y., Nakasato, N., Tanikawa, A., Nomoto, K., Maeda, K., Hachisu, I. 2015, ApJ, 807, 105, doi:10.1088/0004-637X/807/1/105

Shappee, B. J. et al., 2019 ApJ, 870, 13, doi:10.3847/1538-4357/aaec79

Shen, K. J., Boos, S. J., Townsley, D. M. \& Kasen, D. 2021, ApJ, 922, 68, doi:10.3847/1538-4357/ac2304

Siebert, M. R., Dimitriadis, G., Polin, A. \& Foley, R. J. 2021, arXiv:2007.13793

Soker, N., 2011, preprint, arXiv:1109.4652

Soker, N., 2014, MNRAS, 444, L73, doi:10.1093/mnrasl/slu119 

Soker, N., 2018, Sci. Chi. Phys. Mech. Astron., 61, 49502, doi:10.1007/s11433-017-9144-4 

Soker, N., 2019, NewAR, 87, 101535, doi:10.1016/j.newar.2020.101535

Sokoloski, J. L., \& Beldstin, L. 1999, ApJ, 517, 919, doi:10.1086/307234

Sokoloski, J. L., Luna, G. J. M., Mukai, K., \& Kenyon, S. J. 2006, Nature, 442, 276, doi:10.1038/nature04893

Stritzinger, M. D. et al., 2018, ApJL, 864, L35, doi:10.3847/2041-8213/aadd46 

Taddia, F., Stritzinger, M., Phillips, M., et al., 2012, A\&A, 545, L7, doi:10.1051/0004-6361/201220105

Taubenberger, S. 2017, Handbook of Supernovae, 317

Toonen, S., Nelemans, G. \& Portegies Z. S. 2012, A\&A, 546, 70, doi:10.1051/0004-6361/201218966 

Tiwari, V., Graur, O., Fisher, R. et al., 2022,  arXiv:2206.02812

Vallely, P., Fausnaugh, M., Jha, S., et al., 2019, MNRAS, 487, 2372, 
doi:10.1093/mnras/stz1445

van den Heuvel E. P. J., Bhattacharya D., Nomoto K., Rappaport S., 1992, A\&A, 262, 97

Walters, N., Farihi, J., Marsh, T. R. et al., 2021, MNRAS, 503, 3743, 
doi:10.1093/mnras/stab617 

Wang, B \& Han Z., 2010 RAA, 10, 235, doi:10.1088/1674-4527/10/3/005 

Wang, B., \& Han, Z. 2012, NewAR, 56, 122, doi:10.1016/j.newar.2012.04.001 

Wickramasinghe, D. T., Ferrario, L., 2000, PASP, 112, 873, doi:10.1086/316593

Wickramasinghe, D., 2014, Eur. Phys. J. Web Conf., 64, 03001, 
doi:10.1051/epjconf/20136403001

Woosley, S. E. \& Weaver, T. A. 1994, ApJ, 423, 371, doi:10.1086/173813

Yungelson, L., Livio, M., 1998, ApJ, 497, 168, doi:10.1086/305455



\end{document}